\documentclass[prd,aps,preprint,amsmath,nofootinbib,amssymb,eqsecnum,showkeys,tightenlines]{revtex4-1}
\usepackage{epsfig,latexsym,cancel,amssymb,amsmath,verbatim,mathrsfs}
\usepackage{graphicx}
\graphicspath{{figures/}}
\usepackage{subfig}
\usepackage{caption}
\usepackage{subcaption}

\graphicspath{{images/}}
\usepackage{multirow}
\usepackage{float}
\usepackage{slashed}
\usepackage{ulem}
\usepackage[no-math]{fontspec} 
\usepackage{color}

\definecolor{My_red}{cmyk}{0.00,1.00,1.00,0.20}

\usepackage{mathtools}
\usepackage{stix}
\usepackage{float}
\usepackage{xcolor}
\usepackage[colorlinks=true, linkcolor=blue, citecolor=blue, urlcolor=blue]{hyperref}

\begin{document}

\title{A MeV-Scale Dark QCD Solution to the Axion Domain Wall Problem}

\author{Jun Guo}
\email[E-mail: ]{jguo\_dm@jxnu.edu.cn}
\affiliation{College of Physics, Jiangxi Normal University, Nanchang 330022, China}

\author{Zhaofeng Kang}
\email[E-mail: ]{zhaofengkang@gmail.com}
\affiliation{School of physics, Huazhong University of Science and Technology, Wuhan 430074, China}

\author{Jiang Zhu}
\email[E-mail: ]{jackpotzhujiang@gmail.com}
\affiliation{Tsung-Dao Lee Institute and  School of Physics and Astronomy, Shanghai Jiao Tong University,
800 Lisuo Road, Shanghai, 200240 China}
\affiliation{Shanghai Key Laboratory for Particle Physics and Cosmology, 
Key Laboratory for Particle Astrophysics and Cosmology (MOE), 
Shanghai Jiao Tong University, Shanghai 200240, China}

 \begin{abstract}

PQ solution to the strong CP problem probably encounters the axion domain wall problem. In this article, we propose a simple and testable solution, assuming that the $U(1)_{\rm PQ}$ possesses mixed anomaly to a hidden  $SU(N_c)$ color. Then, the axion field receives a new cosine potential from the hidden instantons, which breaks the $Z_{N_{\rm DW}}$ subgroup explicitly. The new potential lifts the vacua degeneracy, but also drives the effective $\theta$ angle away from the origin, re-incuring the strong CP problem. However, we find that the dark QCD scale within the 0.1 to 3 MeV window survives, maintaining a delicate balance. Two observational signatures are explored: gravitational waves from domain wall collapse, already probed by current PTAs, and di-photon signals from axion-dark-glueball mixing, which require next-generation MeV telescopes. The scenario favors a cold dark QCD sector consistent with dark glueball relic constraints.

\end{abstract}


\maketitle

\section{Introduction}

The strong CP problem in the SM is a key window to BSM. The SM allows for a renormalizable term in the QCD sector $\Theta g_s^2/(32\pi^2) G\tilde{G}$, with the gluon field strength tensor $G_{\rho \sigma}^{a}$ and its dual $\tilde{G}^a_{\mu\nu}=\frac{1}{2}\epsilon_{\mu \nu \rho \sigma} G_{\rho \sigma}^{a}$. This term violates parity but is a total derivative term (also known as the boundary term, written as $G\tilde{G}=\partial_\mu K^\mu$) and has no physical consequences in  perturbative calculations. However, the existence of instantons in Yang-Mills theory makes a significant difference.  

Instantons are the gauge field configurations that satisfy the equation $ G_{\mu \nu}^{a} = \pm  \tilde{G}^a_{\mu\nu}$, and are classified by the Pontryagin index $\nu = \frac{g_s^2}{32\pi^2} \int d^4x \, G_{\mu\nu}^a \, \tilde{G}^{a\mu\nu}\in \mathbf{Z}$, which means that the $\theta$-term contributes a nontrivial phase $e^{i\nu\Theta }$ to the path integral. Then, the physical vacuum becomes  the $\theta$-vacuum, a superposition of states with different topological charges, $\lvert \Theta \rangle = \sum_{\nu} e^{i \nu \Theta} \lvert \nu \rangle$. Such vacuum energy receives contribution from the instantons, $\propto \cos \Theta$. In the theta vacuum, the neutron develops EDM, a manifest parity violation. However, the experimental upper limit of the neutron EDM imposes a strong upper limit $\Theta \lesssim 10^{-10}$, causing the puzzle of why it is so small and is called the strong CP problem.

The most promising solution to this puzzle is the Peccei-Quinn (PQ) mechanism~\cite{Peccei:1977ur,Peccei:1977hh}. The key ingredient of the mechanism is a pseudo Goldstone boson $a$, which is produced by the spontaneous breaking of the global $U(1)_{\rm PQ}$ at an energy scale $f_a\gg v_{\rm EW}$; the $U(1)_{\rm PQ}$ is characterized by the mixed anomaly between $U(1)_{\rm PQ}$ and $SU(3)_C$. It generates an anomaly coupling between $a$ and the gluon topological term
\begin{align}\label{aGG}
   \mathcal{A}_3 \frac{a}{f_a} \frac{g_s^2}{32\pi^2}G\tilde{G},
\end{align}  
with $ \mathcal{A}_3$ the $SU(3)_C$ colour anomaly coefficient. In turn, this coupling generates a periodic potential in the form of $\cos (\mathcal{A}_3\theta_{eff})$ with $\theta_{eff}\equiv \Theta/\mathcal{A}_3+a/f_a$. 

From a cosmological perspective, the PQ solution can be subject to the domain wall problem. Specifically, for $ \mathcal{A}_3>1$, the axion potential possesses a $Z_{N_{\rm DW}}$ discrete symmetry, with $N_{\rm DW}=|\mathcal{A}_3|$ known as the domain wall number. As a result, when the axion field takes random values in causally disconnected regions, it randomly rolls down into one of the $N_{\rm DW}$ vacua, leading to the formation of domain walls. If these stable topological defects persist, they would eventually dominate the universe and cause cosmological inconsistency~\cite{Sikivie:1982qv,Sikivie:1982mv}.

However, we note that this issue need not be overemphasized, as models with $ \mathcal{A}_3=1$ avoid it altogether. Even if $ \mathcal{A}_3\neq 1$, the problem can be circumvented cosmologically. If the PQ symmetry breaks before inflation, the entire universe emerges from a single region with a uniform initial value of $\theta$, so the whole universe rolls into the same vacuum among the $N_{\rm DW}$ possibilities, thus evading domain walls. The problem arises only when neither of these conditions is met.

A conventional solution to the domain wall problem is to introduce higher-dimensional operators of the form $\sim |\phi|^{2m}\phi^n/M_{PL}^{2m+n-4}$, which explicitly break the $U(1)_{PQ}$ symmetry and lift the vacuum degeneracy. It has been shown that for a given $f_a(\gtrsim 10^9 \rm GeV)$, only a specific value of $2m+n ~(\geq 9)$ is viable~\cite{Beyer:2023}. In this work, we propose that the axion additionally couples to a dark QCD sector, which dynamically generates the required tilt potential. This mechanism successfully resolves the domain wall problem while preserving the PQ solution to the strong CP problem. The framework favors a relatively cold dark QCD sector, consistent with constraints from dark glueball dark matter relic density. The main observable signatures include gravitational waves from domain wall collapse and di-photon signals from axion–dark-glueball mixing. The gravitational-wave signal is already being probed by current pulsar timing arrays, with the future SKA expected to provide a definitive test. The di-photon signal, while challenging over most of the parameter space, may be accessible to next-generation MeV telescopes in the most optimistic regions.

This article is arranged as the following: in Section II we will first explain the origin of the axion domain wall problem and then propose the dark QCD solution. In Section III we study two most promising phenomenology that may probe such a scheme.

\section{Axion domain wall problem resolved by dark QCD}

\subsection{Quantized Anomaly and the Axion Domain Wall Problem}

First, it is imperative to recognize that the $SU(3)_C$ colour anomaly coefficient  $ \mathcal{A}_3$ must be an integer. The integer nature of $ \mathcal{A}_3$ directly determines the domain-wall number $N_{\rm DW}=|\mathcal{A}_3|$, which counts the degenerate vacua of the low-energy axion potential.  This coefficient originates from the chiral  $U(1)_{\rm PQ}$ anomaly and is formally defined as

\begin{equation}
    \label{}
    \mathcal{A}_3\, 
    = 2 \sum_{q_i}
   Q_{\text{PQ}}^{i}  T(R_i)\in \mathbf{Z} ,
\end{equation}
where the summation runs over all left-handed quarks $q_i$ transforming under the $R_i$ representation of $SU(3)_C$, with $T(R_i)$ denoting the Dynkin index. This integer nature of $ \mathcal{A}_3$ is a direct consequence of the proper normalization of the PQ charges: only relative charge assignments are physically meaningful, and the overall scale can be absorbed into the definition of $f_a$. To see how this works concretely, consider the Kim-Shifman-Vainshtein-Zakharov (KSVZ) model~\cite{Kim:1979if,Shifman:1980zm}.

In the KSVZ model, the complex scalar $S$ carries a PQ charge $q_S$ and is parameterized as $S=(v+\rho)/\sqrt{2}  e^{ia/v }$, where $a$ is the canonical axion field. Coupling $S$ to heavy quarks via $S\bar\Psi_L\Psi_R+h.c.$ yields $f_a=v$ and $\mathcal{A}_3=2q_S T(\Psi)$ in Eq.~\eqref{aGG}. Since the specific value of $q_S$ has no independent effect on the  residual $Z_{N_{\rm DW}}$ symmetry (or on any physical observable), it can be conveniently absorbed into the definition of the decay constant, it can be conveniently absorbed into the definition of the decay constant, i.e., $f_a\equiv v/q_S$. Consequently, for  $T(\Psi)=1/2$---corresponding to a single pair of heavy quarks in the fundamental representation---the KSVZ model serves as a prime example that naturally circumvents the domain wall problem, as outlined in the Introduction. However, in most models, the domain wall number $N_{\rm DW}$ differs from unity. A notable counterexample is the Dine-Fischler-Srednicki-Zhitnitsky (DFSZ) model~\cite{Dine:1981rt,Zhitnitsky:1980tq}, for which $N_{\rm DW}=6$. 

The term \eqref{aGG} represents the explicit breaking of the  $U(1)_{\rm PQ}$ symmetry by the instanton effect, while the discrete subgroup $Z_{N_{\rm DW}}$ remains intact. This can be seen by performing a periodic shift of the axion field: $a \to a+ 2\pi f_a k/N_{\rm DW}  $ ($k=0,1...,N_{\rm DW}-1$), which adds a term $ 2\pi   k  \frac{g_s^2}{32\pi^2}\int d^4x G\tilde{G}=2\pi k\nu$ ($\nu\in \mathbf{Z}$) to the action, thereby leaving the path integral invariant. Below the QCD chiral symmetry breaking scale at $T_{\rm QCD}\simeq 170$ MeV, the non-perturbative dynamics generate an effective potential compatible with this $Z_{N_{\rm DW}}$:
\begin{equation}
    V =- \chi_{\text{QCD}}  \cos\left( \frac{N_{\rm DW} a}{f_{a}} + \Theta \right)=-\chi_{\text{QCD}}  \cos (N_{\rm DW}\theta_{eff}),\qquad \theta_{eff}\equiv \frac{\Theta}{N_{\rm DW}}+\frac{a}{f_a},
    \label{P3}
\end{equation}
where $\chi_{\text{QCD}}\approx (75.6 \rm MeV) ^4$ is the topological susceptibility at zero temperature. Minimising $V$ forces $\langle N_{\rm DW}\theta_{eff}\rangle=0$ in the vacuum, i.e. the effective strong CP phase is dynamically relaxed to zero, solving the strong CP problem. However, for $N_{\rm DW}>1$, it leaves the axion domain wall problem.

When the temperature of the universe drops to the energy scale of QCD chiral symmetry breaking, the QCD instanton effect "turns on" and the $Z_{N_{\rm DW}}$ symmetric axion potential \eqref{P3} emerges. At the same time, the axion field gains a mass from the potential, $m_a^2 = \left. \frac{\partial^2 V}{\partial a^2} \right|_{a = 0} \simeq N_{\rm DW}^2\frac{\chi_{\text{QCD}} }{f_a^2}$ and overcomes Hubble damping, rolling towards one of the $N_{\rm DW}$ vacua $\langle\theta_{eff}\rangle=2\pi k/N_{\rm DW}$ randomly, leading to the formation of axion domain walls interpolating them. These domain walls are characterized by intrinsic tension $\sigma\simeq 9m_a f_a^2 $~\cite{Chang:1998pb}. If they do not collapse before BBN, they come to dominate the energy density of the universe and give rise to the domain wall problem~\cite{Sikivie:1982qv,Sikivie:1982mv}. The present work aims to resolve this problem by invoking a dark QCD sector.

\subsection{Lifting the domain wall degeneracy with a dark QCD sector}

Dark-colored sectors are promising candidates for physics beyond the Standard Model (BSM) and have attracted increasing attention since the discovery of gravitational waves (GWs), as a QCD-like phase transition could be first order and thus provide a GW source in the early universe. A fundamental question concerns the interaction between such dark-colored sectors and the SM sector. One possible approach is to introduce heavy particles within the dark QCD sector that are simultaneously charged under SM gauge groups. These particles generate higher-dimensional operators of the form $\sim \frac{1}{M^4}G_d^2 V_{\rm SM}^2$, where $G_d$ and $V_{\rm SM}$ denote the dark gluons and SM vector bosons, respectively, thereby linking the two sectors. Nevertheless, this approach is largely phenomenological, motivating the exploration of scenarios with deeper theoretical foundations. In particular, the present work demonstrates that an appropriately designed portal between the two sectors can naturally resolve the axion domain wall problem. 

The key observation is that if, analogously to its coupling to visible QCD, the axion also couples to the $SU(N_c)$ dark QCD via the anomalous coupling to dark gluons,
\begin{align}\label{a-G}
   \mathcal{A}_d \frac{a}{f_{a }} \frac{g_d^2}{32\pi^2}G_d\tilde{G}_d.
\end{align}
Consequently, dark instantons also generate a cosine potential for the axion,
\begin{equation}
    V_d = -\chi_0 \cos\left( \frac{\mathcal{A}_d  a}{f_{a}} + \Theta_d \right)=-\chi_0 \cos\left( \mathcal{A}_d  \theta_{eff}+\delta \right),
    \label{dcos}
\end{equation}
where $ \delta\equiv\Theta_d -\mathcal{A}_d/N_{\rm DW}\Theta $, with $\Theta_d$ the dark $\theta$ term of dark QCD, and $\chi_0 $ denotes the zero-temperature topological susceptibility of the dark sector. In general, this potential provides an explicit breaking of the  $Z_{N_{\rm DW}}$ symmetry and therefore lifts the vacuum degeneracy. 
 
At the fundamental level, the effective coupling \eqref{a-G} arises from the mixed $U(1)_{\rm PQ}$ - $SU(N_c)^2$ anomaly. The most economical way to realize this is to assume that the quarks responsible for the PQ anomaly also carry dark $SU(N_c)$ charge. This is a natural option in the KSVZ model, where the relevant quarks are exotic heavy quarks. In contrast, in the DFSZ model the relevant quarks are the Standard Model quarks themselves, which cannot simultaneously carry dark color charges without conflicting with the SM. Hence, one must instead introduce new dark quarks $\psi_d$ coupled to the PQ scalar via $S\bar \psi_d\psi_d$, with $\psi_d$ charged under $SU(N_c)$, analogous to the KSVZ construction.

To determine whether this mechanism is cosmologically viable, however, we should examine the temperature dependence of the dark axion potential. In pure Yang-Mills theory, topological activity is highly sensitive to temperature. The confining phase transition occurs at $T_c\approx (0.6+0.5/N_c^2)
 \sqrt{\sigma_s}$~\cite{Lucini:2003zq}, with $\sigma_s$ the string tension. The intrinsic dark QCD scale satisfies $\Lambda_d\approx 0.5\sqrt{\sigma_s}$, a relation insensitive to $N_c$. In the confined phase ($T<T_c$), topological structures (such as instanton-anti-instanton molecules, calorons) are confined by the non-perturbative vacuum; their size distribution remains essentially set by $\Lambda_d$, so the topological susceptibility $\chi_{\mathrm{top.}}(T)$ is nearly $T$-independent, staying at $\chi_0 \simeq A_0 T_c^4$ with $A_0 \approx 0.2$ and a weak dependence on $N_c$ from continuum-extrapolated lattice data~\cite{Bonanno:2023jbi}. Notably, there is a significant discrepancy between perturbative and non-perturbative calculations in this regime. For example, in pure Yang-Mills theory with $N_c = 3$, the one-loop dilute instanton gas approximation (DIGA) predicts $\chi_{\mathrm{top.}}(T)\approx 4.7T_c^4$, an order of magnitude larger than the lattice result. This highlights the necessity of non-perturbative input for reliable normalization near $T \sim  T_c$.

In the deconfined phase ($T>T_c$), color-electric degrees of freedom are liberated, and the instanton gas becomes highly dilute, with its density suppressed by a factor $\sim e^{-c N_c}$. DIGA gives the following form for the topological susceptibility above $T_c$~\cite{Gross:1981ry}: 
\begin{equation}
    \chi_{\text{top.}}(T) =  \chi_{\text{top.}}(T_c^+) \left(\frac{T}{T_c}\right)^{-n_{\rm YM}}   e^{-S_I(T)+S_I(T_c)},\quad S_I(T)=8\pi^2/g^2(T)\approx \frac{11N_c}{3}\frac{T}{T_c},   
    \label{eq:Tchi}
\end{equation}
where the exponential term originates from the classical instanton action and dominates the temperature dependence at sufficiently high $T$. Lattice simulations~\cite{Bonanno:2023jbi,Bonanno:2026topological} confirm that for $T_c  \lesssim T\lesssim 2 T_c$ the power-law exponent agrees with the DIGA prediction $n_{\rm YM} = 11N_c/3 - 4$. Moreover, they yield $  \chi_{\text{top.}}(T_c^+)/T_c^4\approx 0.45 e^{-0.45 N_c}$. Consequently, the ratio $  \chi_{\text{top.}}(T_c^+)/\chi_0\approx 0.23 e^{-0.45 N_c}$ exhibits a finite jump across the first-order deconfinement phase transition, reflecting the abrupt change in the topological structure of the vacuum.

Armed with this understanding of the temperature-dependent dark axion potential, we now turn to a detailed analysis of whether and how this mechanism resolves the domain wall problem.

\subsection{A precise balance to avoid the domain wall problem} 

Resolving the domain wall problem requires a delicate balance: the dark QCD scale must be low enough to preserve the solution to the strong CP problem, yet high enough to generate sufficient vacuum bias to collapse the walls.

First, we examine whether the effective QCD angle $\theta_{\text{eff}}$ can be driven to a sufficiently small value. It suffices to work in the limit $\chi_{0} \ll \chi_{\text{QCD}}$; otherwise, the two instanton potentials in Eqs.~\eqref{P3} and \eqref{dcos} become comparable, thereby reintroducing the strong CP problem. In this limit, the combined potential drives the effective vacuum angle to
\begin{equation}
  \langle \theta_{eff}\rangle\approx\frac{2k\pi}{N_{\rm DW}}  - \frac{ \chi_0 }{ \chi_{\rm QCD} } \frac{\mathcal{A}_d}{ N_{\rm DW}^2} \left(1-\frac{\mathcal{A}_d^2}{N_{\rm DW}^2}\frac{\chi_0}{\chi_{\rm QCD}}\frac{2k\pi}{N_{\rm DW}} \right)\delta,
    \label{localm}
\end{equation}
which generally deviates slightly from the pure-QCD minimum $2k\pi/N_{\rm DW}$. To ensure $ |N_{\rm DW}\langle \theta_{eff}\rangle-2k\pi |\lesssim 10^{-10}$, it is typically required that
 \begin{align}\label{upper}
 \frac{ \chi_0}{\chi_{\rm QCD}}\simeq \frac{A_0 T_c^4}{ (75.6 \rm MeV) ^4}  \lesssim  \frac{10^{-10}}{\delta} ,
 \end{align}
implying a dark QCD below the MeV scale. Small values of $\Theta$ and $\Theta_d$, and consequently a small $\delta$, help keep $|N_{\rm DW}\langle \theta_{eff}\rangle-2\pi|$ small. This provides the necessary freedom to midly relax the upper bound on the dark QCD scale. For example, for a reasonably small $\delta \sim 10^{-3}$, a critical temperature $T_c\simeq 3$ MeV is allowed. This bound is stringent~\footnote{An exception occurs when $  {\mathcal{A}_d}={N_{\rm DW}}$ and $\Theta=\Theta_d$, which introduces a $Z_2$ symmetry between the two sectors~\cite{Hook:2019qoh}, hence $\delta=0$. In that case, the mirror QCD scale can be much higher, but the axion remains at the origin without requiring a low dark QCD scale.} and independent of the thermal history, serving as a crucial input for the analysis that follows.

The dark cosine potential provides an explicit violation of the $Z_{N_{\rm DW}}$ symmetry, thereby offering a potential solution to the domain wall problem. At the $N_{\rm DW}$ local minimal given in \eqref{localm}, the potential energy reads $ V_k \approx -\chi_{\mathrm{QCD}}  - \chi_{0}  \cos\left( \frac{2\pi \mathcal{A}_d k}{N_{\mathrm{DW}}} + \delta \right)$. Consequently, the original $N_{\rm DW}$ degenerate vacua are split by an amount 
\begin{align}
  \Delta V\sim   V_k-V_{k-1}\approx -2\sin\frac{ \pi \mathcal{A}_d  }{N_{\mathrm{DW}}}\sin\left(\frac{(2k-1) \pi \mathcal{A}_d  }{N_{\mathrm{DW}}}+\delta\right)\chi_0. \label{Vk}
\end{align}
As long as $\mathcal{A}_d/N_{\mathrm{DW}} $ is not an integer, the splitting is non-vanishing and of order $\chi_0$. If this bias is sufficiently large, it can remove the axion domain walls.

In practice, the required splitting is typically very small, with the precise lower bound depending on the cosmological evolution. A rough estimate of the wall collapse time can be obtained as follows~\cite{Sarkar:1995dd}. The vacuum pressure difference $\sim \Delta V$ causes a wall acceleration period with $|a|\simeq \Delta V/\sigma$, where $\sigma=9N_{\rm DW}\chi_{\rm QCD}^{1/2} f_a$ is the wall tension and $\Delta V$ is given by \eqref{Vk}. When this acceleration becomes comparable to the Hubble expansion rate $H(T_{\rm eq})\approx1.66 \sqrt{g_{*,\rm SM}}T_{\rm eq}^2/M_{\rm Pl}$ at a temperature $T_{\rm eq}$, the wall dynamics become dominated by the bias contraction thereafter, and the walls soon collapse within a time much shorter than one Hubble time.

\begin{figure}
    \centering    \includegraphics[width=1.0\linewidth]{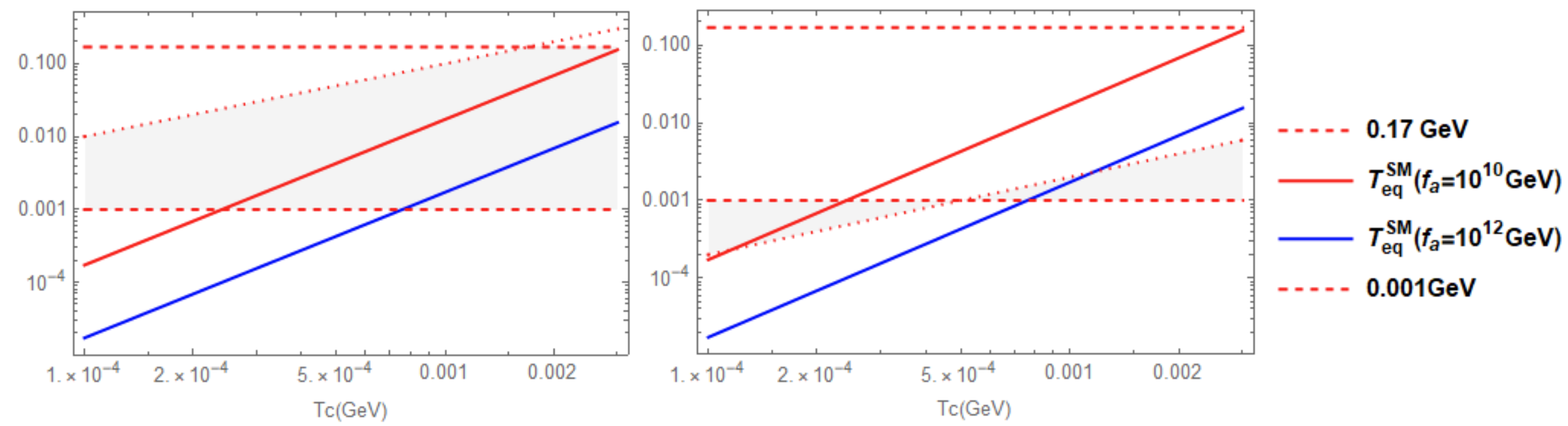}
    \caption{The wall-collapse temperature $T^{\rm SM}_{\rm eq}(T_c)$ (red and blue curves) corresponds to domain walls collapsing after a dark QCD phase transition, which requires it to lie below the dotted red line, $T_c/\xi$. A successful solution to the domain-wall problem further requires $0.001{\rm GeV}<T_{\rm eq}^{\rm SM}<0.17{\rm GeV}$. Left: $\xi=0.01$, right: $\xi=0.5$.}
    \label{collapse:afterTc}
\end{figure}

Let us use the above estimate to derive a lower bound on the dark QCD scale that can resolve the domain wall problem. This requires the collapse to complete before the onset of BBN, which imposes the strong condition $T_{\rm eq}>1\rm MeV$. We first consider the lowest possible of $T_c$, which corresponds to the case $T_{\rm eq}<T_c$~\footnote{This implies that a deconfining phase transition likely occurred before the collapse of the walls; however, the associated gravitational wave signal is too weak to be detected in practice ~\cite{Kang:2021epo,Kang:2024xqk}.}:
\begin{align}\label{Teq:1}
 1{\rm MeV} < T_{\rm eq}\sim \left(\frac{A_0T_c^{4 }}{\Lambda_{\rm QCD}^2}\right)^{\frac{1}{2 }}  \left(\frac{ M_{\rm Pl}}{15 N_{\rm DW}\sqrt{g_{*,\rm SM}}f_a}\right)^{\frac{1}{2 }}<T_c,\quad A_0=0.2. 
\end{align}
In addition, $T_{\rm eq}$ should lie below the QCD phase transition temperature $T_{QCD}$. This window is found to be almost closed due to the upper bound on $f_a$ from the overproduction of the axion by the misalignment of the vaccum $\Omega_a h^2= 0.43\times( f_a/10^{12}{\rm GeV})^{7/6}\langle\theta_{a,i}^2\rangle$, with $\langle\theta_{a,i}^2\rangle\sim \pi^2/3$ the mean square value of the initial angle~\cite{Marsh:2016}. This is illustrated in the right panel of Fig.~\ref{collapse:afterTc}.

\begin{figure}[ht]
    \centering    \includegraphics[width=1\linewidth]{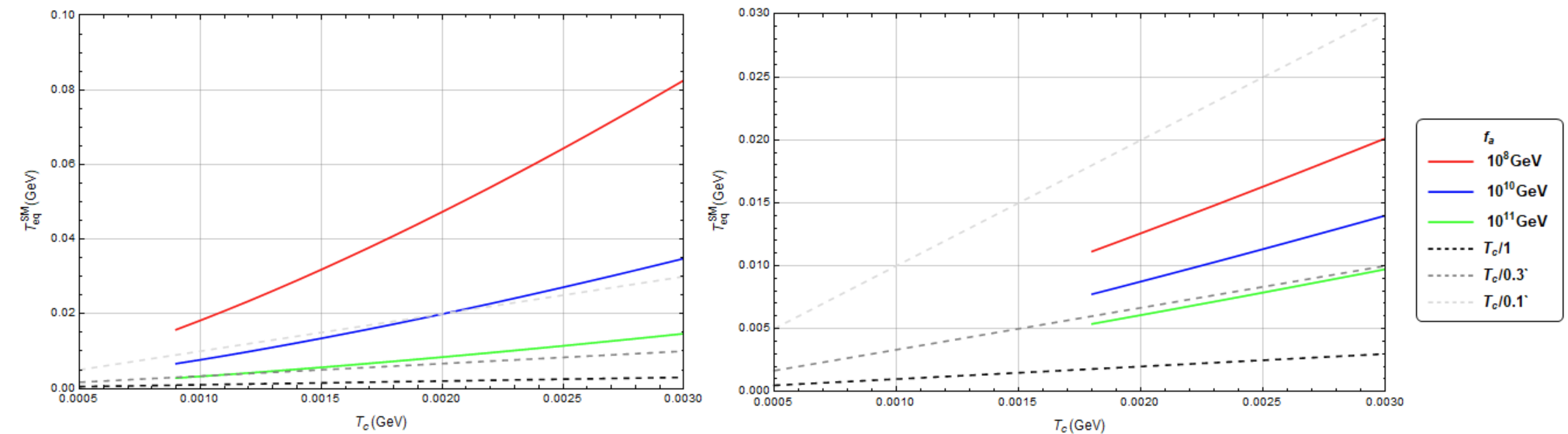}
        \caption{The wall-collapse temperature $T^{\rm SM}_{\rm eq}(T_c)$ corresponds to domain walls collapsing before a dark QCD phase transition, which requires it to lie above the dotted black lines, $T_c/\xi$ for $\xi=1,0.3$ and $0.1$. Left: $N_c=2$; right: $N_c=4$. }
    \label{collape before}
\end{figure}

Instead, consider the opposite case $T_{\rm eq}>T_c$, and replace $\chi_0$ in $V_d$ with with the temperature-dependent susceptibility $ \chi_{\text{top.}}(T)$ to account for the strong dependence of temperature. The same analysis then leads to a wall collapse temperature determined by
\begin{align}\label{Teq:2}
   e^{-S_I(T_{\rm eq}/T_c)} (T_{\rm eq}/T_c)^{-n_{\rm YM}-2}\frac{T_c^2 M_{\rm Pl} }{\chi_{\rm QCD}^{1/2} \sqrt{g_{*,\rm SM}} f_a} \frac{0.23 A_0c_Ve^{-0.45 N_c+11N_c/3}}{15 N_{\rm DW}}\approx 1, 
\end{align}
with $S_I(T)$ given in \eqref{eq:Tchi}. 
The feasible region for $T_c$ is shown in Fig.~\ref{collape before}, confined to a narrow window around 2 MeV. For a fixed $T_c$, the wall collapse temperature decreases as $f_a$ increases. Furthermore, it is an order of magnitude larger than $T_c$, reaching $\mathcal{O}(10)$ MeV or even approaching 100 MeV, which has profound implications for the detectability of this scenario.


So far we have assumed that the dark QCD sector and the SM are in thermal equilibrium, i.e., $\xi\equiv T_d/T_{\mathrm{SM}}=1$. We now relax this assumption and consider a cooler dark QCD with $\xi\ll 1$, which is strongly favored by the dark glueball relic abundance (see Eq.~\eqref{eq:relic}). In this case, the two sectors evolve independently, and the Hubble expansion rate becomes
\begin{equation}
        H(T_{\mathrm{SM}}) \approx 1.66 \sqrt{g_{*,\mathrm{eff}}(T_{\mathrm{SM}})} \cdot \frac{T_{\mathrm{SM}}^{2}}{M_{\mathrm{Pl}}},\quad g_{*,\mathrm{eff}}(T_{\mathrm{SM}}) = g_{*,\mathrm{SM}}(T_{\mathrm{SM}}) + g_{*,\mathrm{dark}}(\xi T_{\mathrm{SM}})\,\xi^{4}, 
    \label{}
\end{equation}
which is well approximated by the SM contribution alone since $\xi\ll 1$. The dark instanton potential $V_d$ depends on the dark gluon temperature $T_d=\xi T_{\rm SM}$. Consequently, the wall collapse conditions derived above must be reinterpreted in terms of the SM temperature $T^{\rm SM}_{\rm eq}$. 

For the scenario where the wall collapses after the dark QCD phase transition (Eq.~\eqref{Teq:1}), the condition becomes 
\begin{align}\label{Teq:1xi}
 1{\rm MeV} < T^{\rm SM}_{\rm eq}\sim \left(\frac{A_0T_c^{4 }}{\Lambda_{\rm QCD}^2}\right)^{\frac{1}{2 }}  \left(\frac{ M_{\rm Pl}}{15 N_{\rm DW}\sqrt{g_{*,\rm eff}}f_a}\right)^{\frac{1}{2 }}<  T_{c}/\xi,\quad A_0=0.2. 
\end{align}
where $T_c$ is the dark QCD critical temperature (now a free parameter), and the upper bound $T_c/\xi$ replaces the earlier $T_c$ because $T^{\rm SM}_{\rm eq}<T_c/\xi$ ensures that the dark QCD sector is already in the confined phase by the time collapse occurs. While this scenario was excluded for $\xi=1$, it becomes viable for $\xi\ll 1$, opening up a broader region in $T_c$. This is demonstrated in Fig.~\ref{collapse:afterTc} for two representative values of $\xi $. For instance, with $\xi =0.01$, $T_c$ can range between 0.1 MeV and 3 MeV。

In contrast, the alternative scenario where the wall collapses before the dark QCD phase transition (Eq.~\eqref{Teq:2}) is disfavored for a cooler dark sector. The right panel of Fig.~\ref{collape before} shows that for $N_c=4$, no viable parameter space exists when $\xi\lesssim 0.1$.

Thus, a cooler dark QCD with $\xi\ll 1$ significantly expands the allowed parameter space for resolving the domain wall problem, primarily through the post-transition collapse channel.

 \section{Two most promising phenomenology}

\subsection{Domain wall collapse and gravitational waves}

The axionic domain wall forms near  $T_{QCD}$ and, after several Hubble times~\cite{Blasi:2025tmn}, the domain wall network enters the scaling region where the energy density satisfies $\rho_{\rm DW}=\mathcal{A} \sigma/t$, with the area parameter $\mathcal{A}\sim 1$. In the radiation-dominated era, $T\propto t^{-1/2}$, so the bias induced by the dark QCD potential becomes significant only much later. Hence, it is justified to neglect the temperature dependence of the wall tension $\sigma$ in the following discussion of wall collapse. 

During the scaling regime, domain walls continuously intersect, reconnect, and accelerate, generating a time-varying quadrupole moment and thus continuous emission of gravitational radiation. However, this source is subdominant compared to the contribution from the wall collapse itself; therefore, we focus on the latter. In general, the GW energy density spectrum at production is defined as $\tilde{\Omega}_{\rm GW}(f)=\frac{1}{\rho_c}\frac{d\rho_{\text{GW}}}{d\ln f}$, with $\rho_c$ the critical energy density $\rho_c\equiv 3H^2/(8\pi G)$. After redshifting to today, the amplitude of this decoupled component is reduced by a factor $(a(T_{\rm eq})/a_0)^4=(g_{*s,0}/g_{*s}(T_{\rm eq}))^{4/3}(T_0/T_{\rm eq})^4$, where $T_{\rm eq}$ denotes the visible-sector temperature at wall collapse (i.e., $T^{\rm SM}_{\rm eq}$ in the previous section), with $g_{*s}(T_{\rm eq})\approx 20$ and $g_{*s,0}=3.91$. The observed GW spectrum then becomes
\begin{equation}
    \label{GWspectrum}
  \tilde{\Omega}_{\text{GW},0} h^{2}=\frac{\rho_c}{\rho_{c,0}}\left(\frac{a(T_{\rm eq})}{a_0}\right)^4\tilde{\Omega}_{\rm GW}(f)h^2
    = \Omega_{r} h^{2} \, \frac{g_{*}(T_{\text{eq}})}{g_{*,0}}
      \left( \frac{g_{*s,0}}{g_{*s}(T_{\text{eq}})} \right)^{4/3}
      \tilde{\Omega}_{\text{GW}},
\end{equation}
where $\rho_{c,0}$ is the present-day critical density, and $\Omega_{r} h^{2} \approx 4.18\times 10^{-5}$ is the radiation energy density fraction (including photons and neutrinos) today. To provide a concrete illustration, we will derive the peak amplitude $\tilde {\Omega}^{\text{peak}}_{\text{GW}}$, which is of primary observational interest.

As a naive estimate, one may assume that during the scaling regime the walls collapse instantaneously at the temperature where the wall tension balances the potential bias~\cite{Hiramatsu:2013qaa}: $\Delta V= H(T_{\text{eq}})\sigma$, which is equivalent to the earlier estimate based on comparing the wall acceleration to the Hubble expansion. At the moment of collapse, the dominant fraction of the domain wall energy is converted into gravitational waves. The intrinsic peak frequency of the GW emission is inversely proportional to this timescale, and its present-day value after cosmological redshift is

\begin{equation}
    \label{}
     {f}_{\text{peak},0} \sim \frac{a(T_{\rm eq} )}{a_0}H(T_{\text{eq}})\simeq 1.3 \, \text{nHz} \left( \frac{g_*(T_{\text{eq}})}{20} \right)^{1/6} \left( \frac{T_{\text{eq}}}{0.01 \, \text{GeV}} \right)
\end{equation}
sector does not affect the result much. 
where the relativistic degrees of freedom are $g_{*\rm SM}\approx 10.75$ before BBN and 3.36 well below it. A dark glueball sector does not significantly affect this result.

Based on the naive quadrupole estimate, the GW energy density at the peak frequency scales as $G\rho_{\rm DW}^2 t^2$, leading to the amplitude
\begin{equation}
\tilde\Omega^{\rm peak}_{\rm GW}
= \epsilon \frac{3}{32\pi}
\left. \left( \frac{2\sigma H}{3H^2 M_{\rm Pl}^2} \right)^{2} \right|_{T_{\rm eq}}
= \frac{\epsilon\ \sigma^{2}}
{24 \times 1.66^2 \pi\ g_*(T_{\rm eq})\ M_{\rm Pl}^{2}\ T_{\rm eq}^{4}}.
\label{eq:GWamplitude}
\end{equation}
Earlier numerical simulations give an efficiency factor $\epsilon\simeq 0.7$~\cite{Hiramatsu:2013qla,Saikawa:2017}; a more refined determination will be discussed below. A lower collapse temperature enhances the amplitude.

Using Eq.~\eqref{GWspectrum}, the present-day GW energy fraction at the peak frequency becomes
\begin{align}
   \tilde {\Omega}^{\text{peak}}_{\text{GW},0}
=
   2.4 \times 10^{-10}\times \, 
    \left(\frac{\epsilon}{0.7}\right)\left(\frac{N_{\rm DW}f_a}{6\times 10^{12}\rm GeV}\right)^2 \left(\frac{1\rm MeV}{T_{\rm eq}}\right)^4\frac{\chi_{\rm QCD}}{\left(0.076\rm GeV\right)^4}.
    \label{eq:GWp}
\end{align}
The upper bound on $f_a$ and the lower bound on $T_{\rm eq}$ set an upper limit on the peak amplitude. When these limiting values are adopted, the predicted signal lies marginally within the sensitivity of current experiments. Near the peak frequency, the GW spectrum is expected to exhibit power-law behavior: the low-frequency tail ($f\ll f_{peak}$) is dominated by the scaling phase, scaling as $\sim f^3$, while the high-frequency tail ($f\gg f_{peak}$) is dominated by the collapse acceleration phase, scaling as $\sim f^{-1}$.

Recent 3+1 lattice simulations, however, reveal significant deviations from these naive estimates~\cite{Notari:2025kqq}. For a domain-wall network with a fixed bias, the simulations show that GW emission persists until a much smaller Hubble scale, $H_{end}\approx 0.1H(T_{\rm eq})$, implying that the actual collapse occurs at a lower temperature $T_{end}= C_s T_{\rm eq}$ with $C_s\approx 0.3$ and enhancing the GW amplitude by roughly two orders of magnitude. Meanwhile, the GW efficiency parameter is found to be $\epsilon\approx 0.07$, which partially offsets this enhancement and reduces the earlier estimate \eqref{eq:GWp}. Furthermore, the resulting GW spectrum exhibits a richer shape; for a periodic potential, it is well described by a doubly broken power-law template,
\begin{equation}
   \tilde \Omega_{\mathrm{GW}} (f)=  \tilde\Omega_{\mathrm{GW}}^{\rm peak}\times S(x) ,
    \quad 
     {S}(x) = 
    \frac{3+\beta + (x_p/x_b)^{\beta+\gamma}}
         {\beta\left(\frac{x}{x_p}\right)^{-3} 
          + 3\left(\frac{x}{x_p}\right)^{\beta} 
          + \left(\frac{x_b}{x_p}\right)^{-\beta}\left(\frac{x}{x_b}\right)^{\gamma}},
    \label{eq:peak_freq}
\end{equation}
where $x \equiv  {f}/{ H_{end}}$. The spectral shape is characterized by several parameters: the peak location $x_p=1.88\pm 0.06$, the high-frequency break point $x_b=10.7\pm 2.3$, the intermediate slope  $\beta=0.61\pm 0.04$, and the high-frequency slope $\gamma=2.34\pm 0.38$ (for $f>f_b$).

When a gravitational wave passes between Earth and a pulsar, it induces periodic stretching and compression of spacetime. This perturbs the light-travel time of the pulsar’s radio signals, producing a tiny advance or delay in their time of arrival---an effect known as a timing residual that varies across the sky. By searching for a specific directional correlation pattern in timing residuals collected over approximately 20 years, pulsar timing array (PTA) experiments have reported strong evidence for a stochastic gravitational-wave background in the nanohertz band ($f \sim 1/\Delta t \sim \text{nHz}$), with $\Omega_{\rm GW}h^2\sim 10^{-10}$~\cite{NANOGrav:2023gor}. With its projected $\mathcal{O}(10^3)$ millisecond pulsars and nanosecond-level residuals, future Square Kilometre Array (SKA) is expected to improve the nHz strain sensitivity by several orders of magnitude (corresponding to a 6-7 order-of-magnitude leap in the $\Omega_{\rm GW}h^2$ detection threshold), offering unprecedented access to the gravitational-wave sky~\cite{Janssen:2014dka}.

\begin{figure}[ht]
    \centering    
   \includegraphics[width=0.32\linewidth]{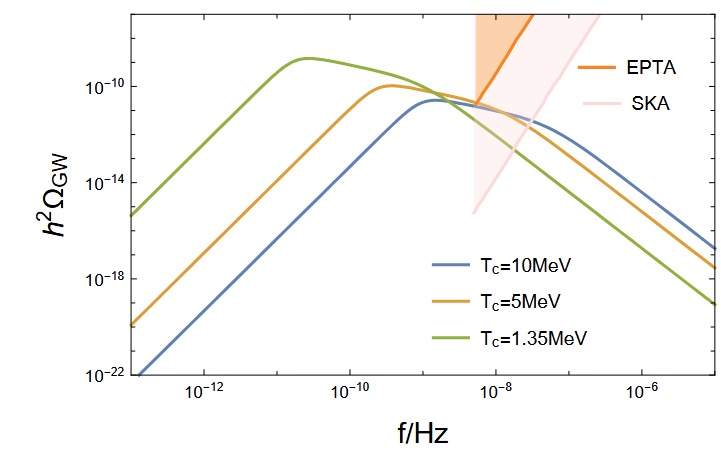}
    \includegraphics[width=0.32\linewidth]{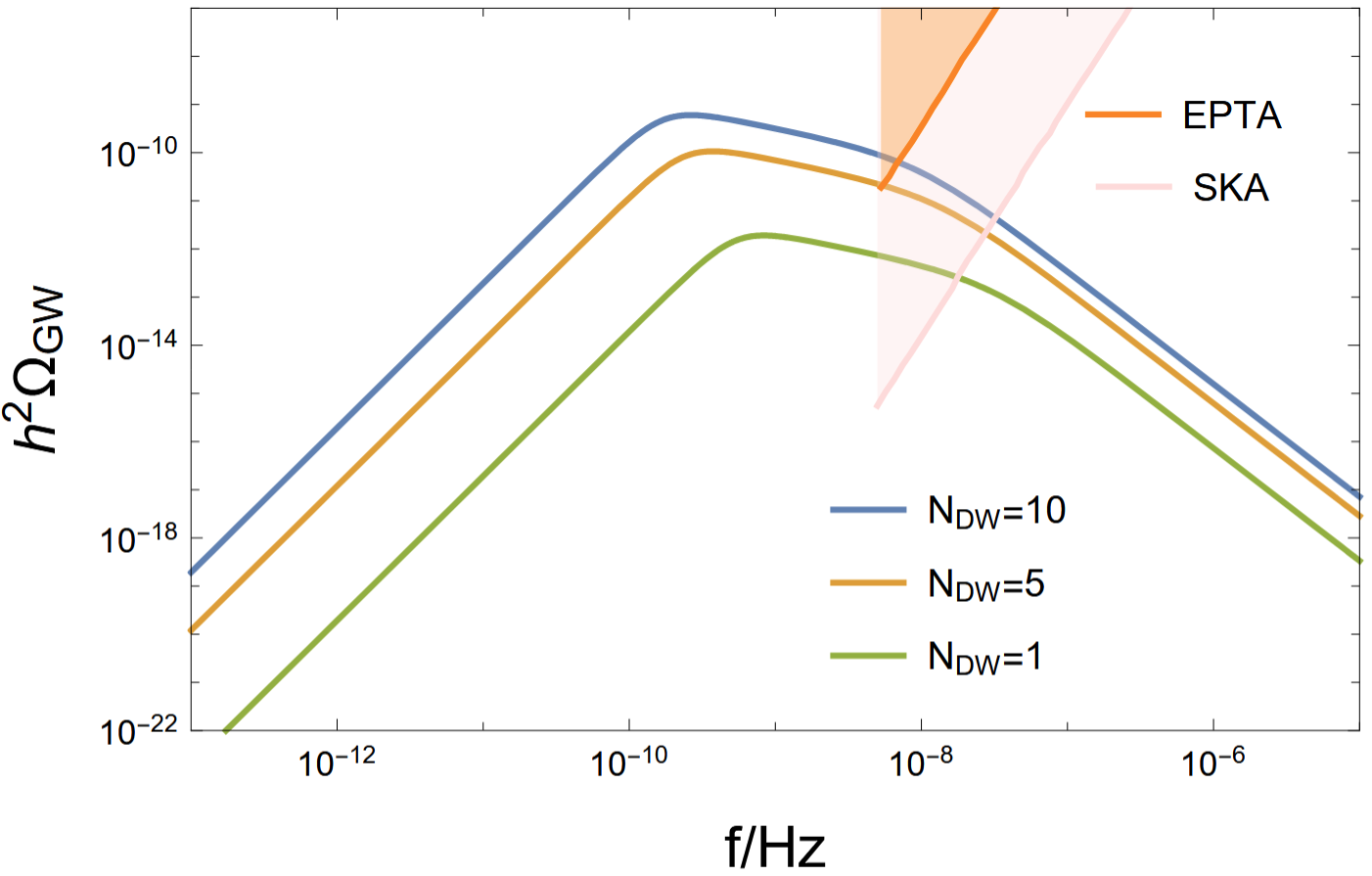}
    \includegraphics[width=0.32\linewidth]{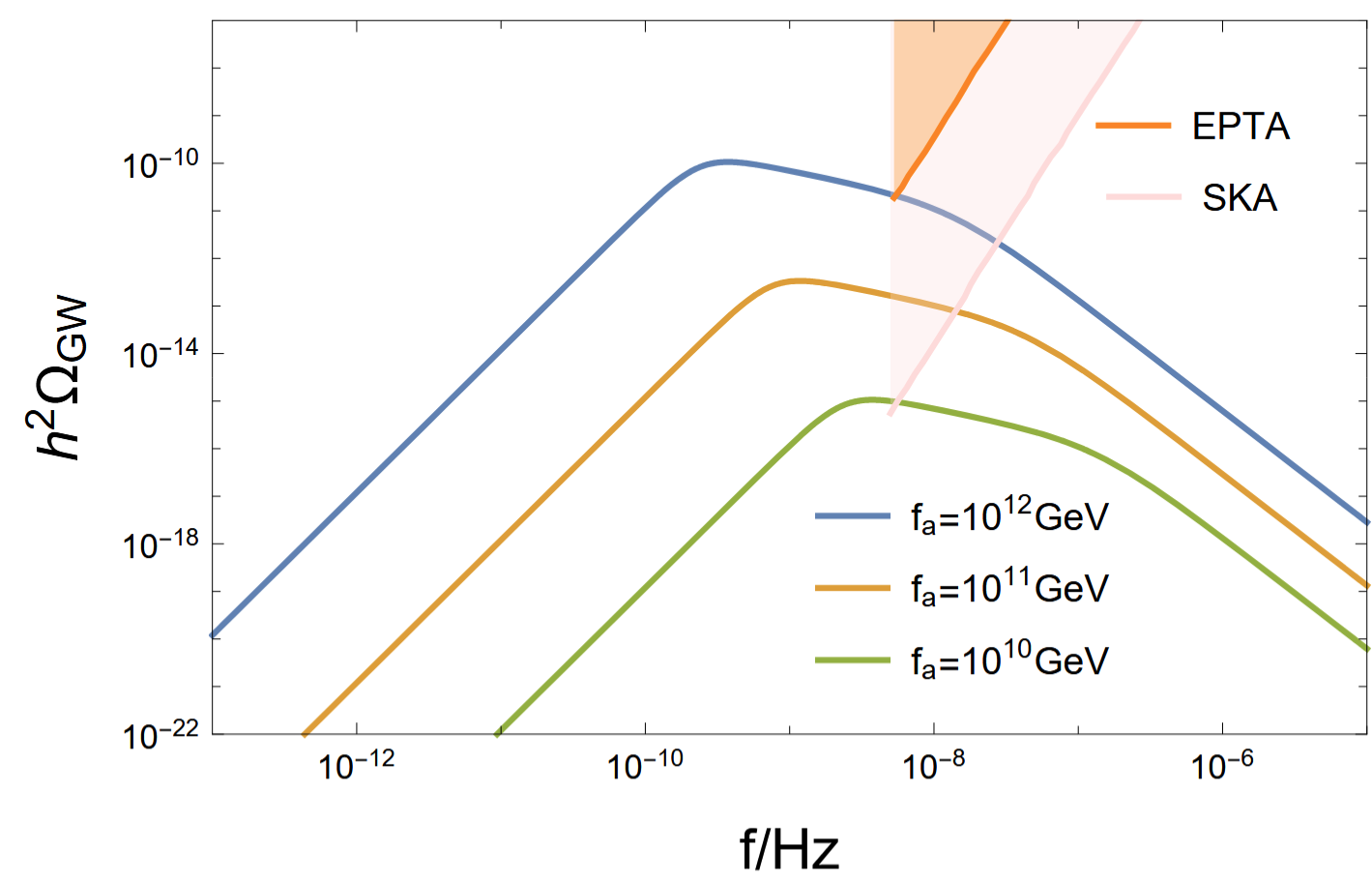}
        \caption{Gravitational-wave spectra from axion domain wall collapse as functions of $T_c$ (left with $N_{\rm DW}=5$, $f_a=10^{12}$ GeV), $N_{\rm DW}$ (middle with $T_c=5$ MeV, $f_a=10^{12}$ GeV), and $f_a$ (right with $N_{\rm DW}=5$, $T_c=5$ MeV), evaluated at the respective benchmark points. The sensitivity bands of current PTAs (orange) and the future SKA (pink) are overlaid for observational comparison.}
    \label{collape GWs}
\end{figure}

Returning to our model, we can use Eq.~\eqref{eq:peak_freq} to predict the gravitational-wave signals from the collapse of axion domain walls. The model involves three phenomenological parameters: $f_a$, $N_{\rm DW}$, and $T_{\rm eq}$, in addition to the fixed input $\chi_{\rm QCD} = (0.076{\rm GeV})^4$ determined by lattice QCD. Through Eq.~\eqref{Teq:1}, $T_{\rm eq}$ can be related to the intrinsic parameter $T_c$. Accordingly, we present the results in terms of $f_a$, $N_{\rm DW}$, and $T_c$, as shown in Fig.~\ref{collape GWs}. The figure illustrates the individual dependence of the GW spectra on $T_c$, $N_{\rm DW}$, and $f_a$.

The parameter dependence can be understood from the dynamics of the domain-wall network. A larger $T_c$ triggers an earlier onset of the bias-induced collapse, when the universe is denser and the Hubble expansion rate is larger. This shifts the characteristic timescale to earlier epochs, leading to higher-frequency gravitational waves, while simultaneously reducing the time available for the network to reach the scaling regime, thereby suppressing the peak amplitude. In contrast, increasing $f_a$ enhances the intrinsic energy stored in the domain walls through their tension, directly boosting the overall gravitational-wave output. Similarly, a larger $N_{\rm DW}$ increases both the wall tension and the number of degenerate vacua, resulting in a denser and more energetic network, and thus a stronger gravitational-wave signal.

Remarkably, for benchmark parameters such as $N_{\rm DW}=5$ and $f_a=10^{12}$ GeV, the predicted peak frequency and amplitude fall within the sensitivity band of current PTAs, as indicated by the orange shaded region in Fig.~\ref{collape GWs} representing the stochastic gravitational-wave background recently reported by NANOGrav and EPTA with $\Omega_{\rm GW}h^2\sim 10^{-10}$~\cite{NANOGrav:2023gor}. This suggests that the model is already being probed by existing observations. Looking ahead, the future SKA will not only extend the sensitivity reach but also enable precise reconstruction of the spectral shape. Because the domain-wall collapse spectrum follows a doubly broken power law (Eq.~\eqref{eq:peak_freq}), distinct from the $f^{2/3}$ scaling expected from supermassive black hole binaries, SKA can discriminate between these sources and potentially provide a smoking-gun signature for the axion domain-wall scenario.

\subsection{Axion-glueballs mixings from low scale dark QCD}

Before discussing the mixing with dark glueballs, we first recall the standard axion-photon coupling, which serves as the foundation for the subsequent phenomenology. The coupling is given by
\begin{equation}
    \mathcal{L} \supset-\frac{1}{4} g_{a\gamma}\, a F_{\mu\nu} \tilde{F}^{\mu\nu} =  g_{a\gamma}\, a \mathbf{E} \cdot \mathbf{B},
    \label{aEB}
\end{equation}
with the coupling strength
\begin{equation}
g_{a\gamma} = \frac{\alpha_{\mathrm{EM}}}{2\pi \left( f_{a} / \mathcal{A}_3 \right)} c_{a\gamma},\qquad     c_{a \gamma} = \frac{\mathcal{A}_E}{\mathcal{A}_3} - \frac{2}{3} \cdot \frac{4 + m_u / m_d}{1 + m_u / m_d}.
    \label{}
\end{equation}
In the KSVZ model, the electromagnetic anomaly coefficient $\mathcal{A}_E$ vanishes since the heavy quarks carry only color charge, yielding $c_{a\gamma}=-1.92$. As we will see, this coupling is inherited by the axion-glueball mixing, thereby opening a di-photon decay channel for the dark glueball and providing a promising observational signature.

The axion portal provides a novel connection to dark QCD, and this article focuses on dark glueballs arising from a pure Yang-Mills sector. Their spectrum is governed by the confinement scale $\Lambda_d $ (or equivalently the string tension  $\sqrt{\sigma_s}$), with masses spanning several times $\Lambda_d \approx 6T_c$ and well-defined quantum numbers $J^{PC}$. The lightest state is the $0^{++}$ glueball, corresponding to the operator $\mathcal{O}_{S}={\rm Tr}G_{\mu\nu} { G^{\mu\nu}}$, with a mass  $m_0\approx (3.37- 3.56)\sqrt{\sigma_s}\approx  7\Lambda_d$ for  $N_c=3$ (reducing to about $6.3\Lambda_d$ in the large-$N$ limit)~\cite{Lucini:2004jg,Athenodorou:2021qvs}, making it a prime dark matter candidate. The effective theory for the $0^{++}$ state includes a term $\sim (0^{++})^5$, which induces a $3\to 2$ number-depletion process. The freeze-out temperature $T_f$ of this process sets  the relic density~\cite{Carlson:1992,Halverson:2017,Forestell:2018}
\begin{equation}
    \Omega_s h^2 \simeq \frac{g_{*s}^d \, \xi^3}{g_{*s}} \, \frac{T_f}{3.6 \, \text{eV}}, \label{eq:relic}
\end{equation}
where $g_{*s}$ and $g_{*s}^d$ denote the entropy degrees of freedom in the visible and dark sectors, respectively. For a pure Yang–Mills dark sector, freeze-out occurs near the confinement scale, so $T_f\simeq \Lambda_d$. Saturating the observed $\Omega h^2\simeq 0.12$ then forces a MeV-scale dark glueball to have $\xi \lesssim 10^{-2}$. For an order-of-magnitude estimate we retain the simple form (\ref{eq:relic}); a more refined treatment using a lattice-fitted glueball effective potential lowers the relic density by about an order of magnitude~\cite{Carenza:2022qwe}, shifting the preferred $\Lambda_d$ or $\xi$ accordingly. Such a small $\xi$ can be achieved, for example, via higher-dimensional connector operators~\cite{Kang:2019izi}.

Nevertheless, in a pure Yang-Mills sector, the  $0^{++}$ glueball is not the only possible dark matter component. The tower of heavier glueballs is also stable, though their abundances are typically much smaller. In particular, the parity-odd  $0^{-+}$ state has a slightly higher mass, $m_{0^{-+}}\approx 1.5m_0$, so its relic density is suppressed by roughly a factor $\sim (m_{0^{-+}}/m_{0^{++}})^{3/2}    e^{-(m_{0^{-+}}-m_{0^{++}})/T_f}\approx 0.06$. These two states can mix in the presence of a dark $\theta$-term. Their mixing with the axion will be discussed in the following. 

The topological term \eqref{a-G} induces a direct mixing between the axion and the pseudoscalar dark glueball $0^{-+}$ via the dimension-four operator $\mathcal{O}_{PS}={\rm Tr}G\tilde{ G}$, with the decay constant defined by $\langle 0 | \mathcal{O}_{PS} | 0^{-+} \rangle = F_{0^{-+}}^{P}$. For SU(3) pure Yang-Mills, lattice calculations give the scalar glueball matrix element $4 \pi \alpha_d F^{S}_{0^{++}} = 3.06\, m_0^{3}$ with $\alpha_d\equiv g_d^2/4\pi$ and, in the same work, the pseudoscalar counterpart $4\pi \alpha_d F_{0-+}^{P} =   m_0^3$~\cite{Chen:2005mg}; extending these $SU(3)$ results to a general dark gauge group $SU(N_c)$ relies on large-$N_c$ scaling, so we shall therefore adopt $4\pi \alpha_d F_{0-+}^{P} = c_P\, m_0^3$ with $c_P\approx 0.83$, where $m_0$ is the lightest scalar glueball mass in the given $SU(N_c)$ theory.

The dark glueball can decay into a pair of photons through a virtual axion, $0^{-+}\to a^*\to 2\gamma$~\footnote{For an alternative scenario where parity violation in the dark QCD itself generates an axion-like glueball without a genuine axion, see Ref.~\cite{Carenza:2024qaq}.}. This proceeds via the matrix element $\langle 2\gamma| a |0\rangle  \frac{1}{m_P^2-m_a^2} \langle0|\mathcal{O}_{PS} | 0^{-+} \rangle $, leading to the decay width
\begin{equation}
    \label{}
    \Gamma_{0^{-+} \to 2\gamma}
    =
    \left(
        \frac{\mathcal{A}_d \alpha_d F_{0^{-+}}^{P} }
             {8 \pi f_{a} \left(m_{P}^{2} - m_{a}^{2}\right)}
    \right)^{2}
    \Gamma_{a \to \gamma\gamma} \!\left(m_{P}\right)\equiv \sin^2\beta 
    \Gamma_{a \to \gamma\gamma} \!\left(m_{P}\right), 
\end{equation}
where $\beta$ is the effective mixing angle between the axion and the glueball, and the axion decay width is
\begin{equation}
    \Gamma_{a \to \gamma\gamma}(m_P) = \frac{m_P^3 g_{a \gamma}^2}{64\pi}.
    \label{}
\end{equation}
The dark glueball's coupling to photons is highly suppressed by the mixing angle $\beta\sim \frac{1}{32\pi^2}m_0/f_a\lesssim 10^{-12}$, rendering all signatures of dark glueballs vanishingly small. Consequently, conventional search strategies for axion-like particles have no chance of detecting it. Explicitly, the width of the dark glueball decaying into a photon pair is
\begin{equation}
    \label{lifetime}
    \Gamma_{0^{-+} \to 2\gamma}\approx \frac{\mathcal{A}_d^2c_P^2\alpha_{\rm EM}^2c^2_{a\gamma}}{2^{20}\pi^6}\frac{m_0^5}{ f_a^4}\Rightarrow \tau_{0^{-+}}\approx 10^{31}\times \left( \frac{m_P}{10\rm MeV} \right)^{-5}\left( \frac{10^8 \rm GeV}{f_a} \right)^{-4}s,
\end{equation}
In the above estimate, we have set $c_{a\gamma}=1$ and $\mathcal{A}_d=1$. If these parameters are substantially enhanced, the lifetime can be reduced by up to two orders of magnitude. Furthermore, the upper bound in Eq.~\eqref{upper} imposes a maximal mass for the dark glueball, $m_P\approx 1.5\times 6T_c\lesssim 30$ MeV. For such a heavy state, the lifetime could be shortened by an additional two orders of magnitude.

Before turning to experimental sensitivities, we note an additional feature: the dark $\theta$-term also mixes the $0^{-+}$ with $0^{++}$ states. Consequently, the $m_{0^{++}}$ glueball inherits a coupling to photons through this mixing, giving rise to a second gamma-ray line at $m_{0^{++}}/2$. The strength of this line depends on the $0^{-+}$ with $0^{++}$ mixing angle, which is controlled by the dark $\theta$-term. If the mixing is too small, the signal is suppressed by a factor  $\sim \Theta^2$~\footnote{This is a very rough estimate; a reliable determination of this mixing would require non-perturbative methods.} and becomes undetectable. However, if the mixing is sizable, the resulting double-line structure ($m_{0^{++}}/2$ and $m_{0^{-+}}/2$) would provide a distinctive signature of the scenario proposed here.

Existing observations sample the lower part of our glueball mass range but do not yet impose meaningful constraints. The high-resolution spectrometer SPI on board INTEGRAL provides some of the strongest published limits on narrow photon lines below a few MeV. Based on nearly 20 years of INTEGRAL/SPI data, Ref.~\cite{Fischer:2022pse} searched for line-like signals from decaying dark matter in the mass range $40\,{\rm keV}<M_{\rm DM}<14\,{\rm MeV}$, reporting limits of order $\tau_{\rm DM}^{\rm SPI}\sim1\times10^{26}-1\times10^{27}\,{\rm s}$. A complementary analysis using 16 years of INTEGRAL/SPI data with a dark-matter spatial template~\cite{Calore:2022pks} extends the reach to $\tau_{\rm DM}^{\rm SPI}\sim1\times10^{27}-1\times10^{29}\,{\rm s}$ for $60\,{\rm keV}\lesssim M_{\rm DM}\lesssim16\,{\rm MeV}$. For a two-photon decay, these bounds apply mainly to photon energies below approximately  $7-8\,{\rm MeV}$, i.e., below the endpoint  $m_P\simeq30\,{\rm MeV}$  of our glueball spectrum. Even in the most optimistic corner of our parameter space---with enhanced couplings $c_{a\gamma}$ and $\mathcal{A}_d$ and a glueball mass near the upper bound $m_P\simeq30\,{\rm MeV}$—the predicted lifetime $\tau_{0^{-+}}\sim 10^{29}s$ barely reaches the upper edge of the SPI sensitivity window, while the corresponding photon energy $E_\gamma\simeq 15\rm MeV$ lies outside the SPI energy coverage.

Consequently, existing INTEGRAL/SPI data do not constrain our model, but they establish a useful reference for assessing the prospects of next-generation missions. Among these, COSI targets the low-mass part of our glueball spectrum, while next-generation MeV telescopes such as AMEGO-X and e-ASTROGAM provide complementary coverage of the high-mass part:
\begin{itemize}
\item \textbf{COSI}~\cite{Tomsick:2019wvo}: a soft $\gamma$-ray Compton telescope covering
$0.2\ {\rm MeV}\lesssim E_\gamma\lesssim5\ {\rm MeV}$, optimized for spectral-line searches. For a diphoton decay this translates to $0.4\ {\rm MeV}\lesssim m_P\lesssim10\ {\rm MeV}$. Dedicated forecasts give $\tau_{\rm DM}^{\rm COSI}\sim10^{27}{-}10^{29}\ {\rm s}$~\cite{Caputo:2022dkz}, making COSI most relevant for the {lower} part of our glueball spectrum, but not for the endpoint $m_P\simeq30\ {\rm MeV}$.
\item \textbf{AMEGO-X and e-ASTROGAM}~\cite{Caputo:2022xpx,e-ASTROGAM:2017pxr}: these next-generation MeV missions (from NASA and ESA, respectively) cover the energy range $100\ {\rm keV}\lesssim E_\gamma\lesssim1\ {\rm GeV}$ (AMEGO-X) and $0.3\ {\rm MeV}\lesssim E_\gamma\lesssim3\ {\rm GeV}$ (e-ASTROGAM), with projected line sensitivities of $\tau_{\rm DM} \sim10^{28}{-}10^{30}\ {\rm s}$ over the MeV to tens-of-MeV range~\cite{ODonnell:2024aaw}. Both missions are capable of probing the high-mass part of our glueball spectrum, particularly the region $E_\gamma\lesssim15\ {\rm MeV}$ ($m_P\lesssim30\ {\rm MeV}$), and together serve as complementary NASA/ESA benchmarks for the upper end of our mass window.
\end{itemize}
 
In summary, while the non-resonant axion-glueball mixing suppresses the line flux over most of the parameter space, the most optimistic corners---those with enhanced mixing parameters or a glueball mass near the upper bound $m_P\lesssim30\ {\rm MeV}$---may lie within the reach of next-generation MeV telescopes such as COSI, AMEGO-X, and e-ASTROGAM. The possible double-line structure, if confirmed, would provide a powerful discriminant for this scenario.

\section{Conclusion and discussion} 

We have presented a dark QCD solution to the axion domain wall problem, in which a mixed anomaly with a hidden $SU(N_c)$ generates a dark instanton potential that breaks the $Z_{N_{\rm DW}}$ symmetry while keeping the dark QCD scale within a safe MeV window. Rather than attempting a comprehensive study of the full dark matter sector---which in our framework consists of both axions and dark glueballs---we have focused on identifying the most promising observational signatures. Unlike the parity-violation scenario of Ref.~\cite{Carenza:2024qaq}, our setup features a genuine axion that mixes with the pseudoscalar glueball via the topological anomaly, leading to a distinct phenomenology rooted in the PQ mechanism itself. 

A noteworthy feature of our scenario is the complementary behavior of the two main observables. The gravitational-wave signal from domain wall collapse and the di-photon line from glueball decay respond oppositely to changes in the axion decay constant $f_a$ and the dark QCD critical temperature $T_c$ (compare Eqs.~\eqref{eq:GWp} and \eqref{lifetime}, with $T_{\rm eq}$ replaced by $T_c$). This compensation implies that if both detection channels reach sufficient sensitivity, the scenario can be tested in a robust, cross-checking manner.

Regarding the observational prospects for the dark glueball, the situation is challenging. The non-resonant axion-glueball mixing suppresses the line flux severely, with lifetimes given by Eq.~\eqref{lifetime}. Existing gamma-ray data do not constrain our parameter space, and future MeV telescopes can only access the most optimistic corners—those with enhanced couplings ($c_{a\gamma}, \mathcal{A}_d$) or a glueball mass near the upper bound $m_P\simeq30\,{\rm MeV}$. Over most of the parameter space, an improvement of several orders of magnitude in sensitivity would be required. Furthermore, if the pseudoscalar glueball constitutes only a fraction of the total dark matter, the indirect-detection constraints become even weaker.

In summary, while the dark QCD solution elegantly resolves the domain wall problem while preserving the PQ mechanism, its observational verification---especially through di-photon signals---demands significant advances in experimental capabilities. The gravitational-wave channel, on the other hand, already touches the parameter region accessible to current PTAs, offering a more immediate path to testing the idea.

\noindent {\bf{Acknowledgements}} 

This work was supported by the National Natural Science Foundation of China under Grant No. 12305111, Jiangxi Provincial Natural Science Foundation 20252BAC200168

\bibliographystyle{unsrt}  
\bibliography{theta}

\end{document}